# Challenges of Reliability Assessment and Enhancement in Autonomous Systems

Maksim Jenihhin[1], Matteo Sonza Reorda[2], Aneesh Balakrishnan[1,3], Dan Alexandrescu[3]

[1]*Tallinn University of Technology, Estonia, maksim.jenihhin@taltech.ee*
[2]*Politecnico di Torino, Italy, matteo.sonzareorda@polito.it*
[3]*IROC Technologies, France, { aneesh.balakrishnan |dan.alexandrescu }@iroctech.com*

*Abstract*— The gigantic complexity and heterogeneity of today's advanced cyber-physical systems and systems of systems is multiplied by the use of avant-garde computing architectures to employ artificial intelligence based autonomy in the system. Here, the overall system's reliability comes along with requirements for fail-safe, fail-operational modes specific to the target applications of the autonomous system and adopted HW architectures. The paper makes an overview of reliability challenges for intelligence implementation in autonomous systems enabled by HW backbones such as neuromorphic architectures, approximate computing architectures, GPUs, tensor processing units (TPUs) and SoC FPGAs.

## I. Introduction

Recent rapid expansion of autonomous systems has enabled numerous unprecedented novel services and businesses. However, the unleashed benefits come along with computationally extremely challenging mission- and safety-critical application scenarios. The gigantic complexity and heterogeneity of today's advanced cyber-physical systems and systems of systems is multiplied by the use of avant-garde computing architectures to employ artificial intelligence based autonomy in the system. The setups such as swarms of autonomous robotic vehicles are already on the doorstep and call for novel intelligent approaches for reliability that are often the key enabling factor for a new product or technology on the way to market. This success is supported by the connectivity solutions being developed in the IoT research discipline that is also moving towards enhanced autonomy of the connected intelligence-enabled things [1].

Expectations for reliability are very wide as also the variety of autonomous systems. The latter are driven by a number of killer applications listed below:

- *autonomous vehicles* in the automotive domain is the dominant application in terms of funding and recent research efforts invested, includes cars with Autonomous Driving (AD) through levels 3 to 5 of autonomy;
- *aircrafts* with different degree of autonomy, e.g. employing the 'fly-by-wire' reliability-critical systems;
- *unmanned aerial vehicles* (UAVs) commonly known as drones, both fixed-wing and rotary (quadcopters), these days are equipped with very high degree of intelligence and tend to operate autonomously individually or in *UAV swarms* (autonomous System of systems);
- *unmanned ground vehicles* (UGVs) include among others rapidly developing *self-driving delivery robots* (e.g. [2]) and *farming robots*;
- *unmanned underwater vehicles* (UUVs), e.g. robotic fishes, and *unmanned boats* (also *unmanned surface vessel* (USV)) are heading at long-term operation in harsh environments;
- *autonomous spacecrafts* such as satellites and autonomous landers for remote missions often with limited communication capabilities;
- *autonomous military and law enforcement applications*, e.g. that may be dual use of the above mentioned systems but also specific weapons, e.g. autonomous missiles.

Today, autonomous systems are quickly getting on top of the hype cycle [3] and several very recent studies have started to look into the enabling aspects of such systems, e.g. from the standards perspective [4], from the security perspective [5], or for a specific application [6]. In this paper, we make an overview of reliability challenges in autonomous systems. The rest of this paper is organized as follows. Section II outlines the challenging attributes of autonomous systems. Section III and IV target at understanding industrial standards and the key concepts in reliability and safety for autonomous systems Sections V and VI analyze specific requirements introduced by novel applications and architectures and Section VII discusses reliability enhancement. Section VIII wraps up the paper.

## II. Autonomous Systems' Attributes from the Reliability Perspective

The attributes of an *autonomous system* (AS) from the reliability requirements perspective may be summarized in the following set of challenges:

*External attributes:*
a. Specific application domains and operating environments, i.e. dangerous, tedious, remote/hardly- or in-accessible for human involvement;
b. Limited availability and latency of external support for repair or critical decision-making;
c. Real-time constraints (often hard real time);
d. An AS is usually a cyber-physical system immersed into physical world through intensive interaction by sensors (also implying sensor fusion) and actuators [7];

e. Often, several ASs are combined into a System of Systems (SoS) [8] with intensive machine-to-machine communication and resource sharing, enabling computing continuum and complex distributed computing architectures, e.g. edge-to-fog computing;
f. An AS imply subjectively higher expectations to reliability level and lower tolerance to unsafe behavior compared to a human-operated system.

*Internal attributes:*

g. High complexity of the computing architectures capable to run computation-intensive evolvable artificial intelligence software (e.g. the novel GPUs, the TPU for Google's TensorFlow and similar);
h. Specificity of Hardware Neural Networks implementations that are rather a "sea of elements", with reduced structural/functional modularization of hardware;
i. ASs are built utilizing a combination of many very new untested in-field technologies;
j. An AS implementation has strong dependency on the quality of assumptions about the ambient, often dynamic and uncertain environment.

The state-of-the-art academic solutions, e.g. [9],[10],[11], are either incapable or inefficient to tackle this union of challenges. The practical reliability drivers in today's designs are industrial standards in different application domains such as [12] and its application-specific derivatives, e.g. [13], that do not address cross-layer approaches. The standards mostly address functional safety of the complete system rather than just a component and depend on the integrated operation of all sensors, actuators, control devices, and other units. Therefore, the functional safety standards are usually unspecific to the solutions at the integrated circuit (IC) level. The standard for IC stress test [14] in the automotive domain covers requirements for a subset of reliability issues (such as NBTI, HCI, electromigration, etc.) at the chip level. However, many recent application domains unleashed by the unmanned systems, e.g. UAVs, remain uncovered [15],[16].

## III. UNDERSTANDING RELIABILITY STANDARDS IN AUTONOMOUS SYSTEMS

Autonomous systems will enable huge societal changes (and possibly progress). As expected, stringent safety and reliability expectations and requirements are firmly set in international standards, implicit customer expectations and, not unexpectedly, insurance policies. Autonomous systems are also an emerging industrial field and are very likely to stay with us for a very long time. Accordingly, it is very probable that many successive, evolutionary or revolutionary standards will be issued to govern them. International standards are the clearest and most authoritative prescribers regarding reliability and safety. The list of current or under-development standards in this field includes:

- IEC 61508 [12] (Functional Safety of Electrical/Electronic/Programmable Electronic Safety-related Systems) is aimed at all industrial fields and is the template for many application-specific standards;
- One of the most well-known derivatives of the previous is ISO 26262 [13], which addresses the functional safety of automotive systems;
- IEC 62279 is an adaptation of [12] for railway applications;
- ISO 13849 [17] is a safety standard which applies to parts of machinery control systems that are assigned to provide safety functions;
- AC 25.1309-1A [18] (System Design and Analysis) provides background for important concepts and issues within airplane system design and analysis;
- RTCA/DO-254 [19] (Design Assurance Guidance for Airborne Electronic Hardware) provides guidance for the development of airborne electronic hardware.

Since change is a permanent feature of the industrial progress, expectations and requirements constantly evolve. While intended to be robust and durable, standards are not safe from being prone to latest fashions and currents in the industry or from being influenced by companies and organizations looking to promote their own position and offering.

Particularly, the terminology and dictionary of any standard is a faithful snapshot of the particular context at the time of the writing and often suffers from updates, changes of signification, meaning overcharges and obsoleteness during the expected lifetime of a standard and even more so when a new standard is devised. The goal of this Section is to pinpoint some basic topics that are common to the different standards and faced by most of them. They are summarized in Table I.

Many standards include a part related to *Terminology*. In this category, the signification and a clear definition of the key terms shall be presented and elaborated. However, the specific meaning can hide behind an ordinary word, requiring a more in-depth discussion and explanation and investing the simple term with a fundamental weight. The "Terminology" category would thus benefit from a *Concepts* sub-category.

As soon as the key terms and concepts have been introduced, the standards are fast to move to the explanation of their core methodology, framework and principles. The *Methodology* category covers these aspects. In their various proposed methodologies, many of the standards address "risks" to the safety of the intended applications and set a mix of quantitative and qualitative requirements and expectations for these risks. These objectives and goals will be captured in the *Requirements* category. The reliability and safety of any application will have to be checked against the applicable requirements and improved until its behavior fulfils the expectations of the intended standard. Accordingly, the taxonomy will have to include the *Assessment* and *Management* categories.

TABLE I. MAIN TOPICS IN THE RELIABILITY AND FUNCTIONAL SAFETY STANDARDS

| Terminology | Methodology | Requirements | Assessment | Management | Environment |
|---|---|---|---|---|---|
| Vocabulary | Development: | Hazard & Risks | Models | Online | Electrical |
| Concepts | System-level | Classification | Probabilistic | Offline | Thermal |
| | Hardware-level | Event rates | Simulation | Diagnostic | Mechanical |
| | Software-level | Mitigation | | Maintenance | Radiation |

Lastly, any application is designed to work safely and reliably in a given setting. The *Environment* category would capture the entirety of electrical, thermal, mechanical, radiative conditions to which the application will be subjected.

In practice, the concerns about reliability may mix together with those about feasibility (especially when target features are particularly challenging). For this reason, independently on standards and regulations, general safety praxis can be utilized e.g. by using the ALARP method ("as low as reasonably practicable") and providing justification for benefits of the society against the involved risks.

## IV. Key Concepts in Reliability and Functional Safety for Autonomous Systems

For autonomous systems, but not only, the notions of "*reliability*" and "*safety*" comprise as many significations as engineers from different industries want to invest in them. Loosely, reliability represents the probability of a system to fail, i.e. higher reliability means less failures, while safety generally means that the system fails in a safe way. A reliable system can be unsafe while a safe system can be unreliable. Furthermore, systems can be made arbitrarily safe and reliable with a corresponding investment of resources and time. Requirements for reliability and safety can be quantitatively and qualitatively very different but standards are often aggressive in setting high requirements for both safety and reliability. The most straightforward approach to address both reliability and safety is to rank risks and hazards according to their impact (*safety*) and to expect that the probability of risks (*reliability*) decreases inversely to their impact. An aggregated event rate (often measured in terms of Failure in Time, or FIT), may be associated to the system and/or component according to their role but with an underlying understanding of the risks that make up the "Failure" key term.

In this way, quantity and quality, safety and reliability are harmoniously integrated. However, reliability engineers will find that this task is relatively difficult as two opposing concepts still need to be conciliated: objective versus subjective. The qualificative of "*Objective*" can be applied to any physical measurements. As an example, technology fault rates can be expressed accurately; a "*Soft Error Rate*" is an objective measurement of the susceptibility of a technological process under radiations. Faults propagate through the circuit and system and can become Failures. Various methods, such as static and dynamic ones, can accurately and undisputedly (thus objectively) predict the fact that a fault occurring in a deeply-embedded logic cell instance can propagate and affect a primary system output. The question that the reliability engineers and their design colleagues must answer now is whether this fault consequence represents a failure or not, what are the actual consequences and, more importantly, where exactly in terms of risk levels the failure needs to be classified. This is the "*Subjective*" part and standards try to address this by a prescriptive, function-based assessment. However, in practice, the whole procedure provides some freedom and margins to reliability engineers that can argue for a less critical classification of possible fault outcomes.

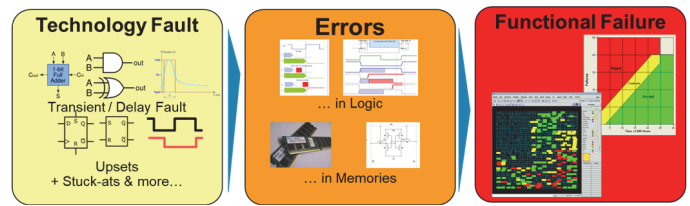

Fig. 1. Faults, errors and failures in a system

Probabilistic risk evaluation and management is a core concept of many reliability assessments. Only a fraction of technological faults will propagate through the circuit and become errors, i.e., erroneous data or values stored instead of correct information. Only a percentage of errors will become failures causing observable deviations of the system behaviour. Furthermore, failures can be classified in criticality classes. If error detection/correction/management features are implemented, they can address faults, errors and failures at any design level and can reduce the percentage of events graduating from one level to the upper one (see Fig. 1).

A first, fundamental contributor to the quality of an autonomous system is the *quality of the underlying implementation technology*. The manufacturing process must present a well-characterized, preferably low intrinsic defect and fault rate, resiliency to environmental challenges and a good, well known aging and degradation performance. Moreover, the technology providers (foundries) must offer their customers a full ecosystem with the tools, IPs and solutions for reliable and safe circuit design.

A second contributor lies in integrating into the system some solutions for *lifetime performance assurance*. The classical bathtub curve is no longer an evidence and the reliability of the system must be managed during the expected lifetime through online and offline monitoring, embedded sensors, test instruments and safety mechanisms (see Fig. 2).

*Configurability and Adaptability*, to environment and workload challenges, as well as to intrinsic degradation and aging, are important for today's autonomous systems running dynamic applications in diverse environments.

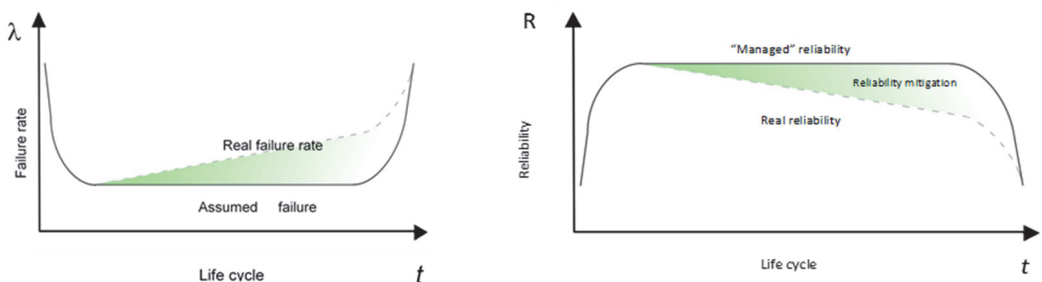

Fig. 2. Managed lifetime reliability (courtesy of the RESIST project).

Lastly, the evolution to "*Self-*" Everything (self-monitoring, self-calibration, self-adaptation, self-configuration, etc.) is an important industry trend and goal that can provide solutions for more reliable and safer autonomous systems.

## V. NOVEL AUTONOMOUS SYSTEMS' APPLICATION-SPECIFIC REQUIREMENTS FOR RELIABILITY VALIDATION

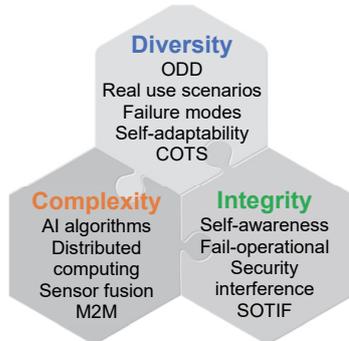

Fig 3. The three clusters of application requirements in autonomous systems

The plethora of new applications unleashed by autonomous systems introduce novel or bring to the front existing requirements for the system reliability validation. These can be represented in the following three clusters as shown in Fig. 3.

### A. Diversity

One of the main challenges for the autonomy is the diversity of the environment and operational conditions. A solution to simplify the problem is to split it to a limited (countable) number of *Operational Design Domains* (ODD), e.g. as introduced by the NHTSA agency. Considering a particular autonomous system, these may include such factors as operational terrain, environmental and weather conditions, communication modes, etc. [23].

Proper reliability validation implies capturing *real world scenarios* along with *real failure modes* that may imply a significant amount of statistical data collected. For example, the calculations in [24] demonstrate that this results in billions of miles of in-field test. This approach is also valid for any-scale safety-critical autonomous systems, e.g. UGVs [2].

The implementation of a complex autonomous system or even a system of systems may involve a *diversity of available fault tolerance structures* in the components employed, e.g. involving COTS parts along with hardened ones. Moreover, as mentioned in the above section, the diversity of autonomous systems is amplified by their features, such *self-adaptability* and *self-configuration*.

### B. Complexity

Autonomous systems have introduced a conceptually new level of complexity for reliability validation. Here, application of *artificial intelligence algorithms* is the main factor for the rapid increase of complexity in HW architectures (see Section VI). This includes in particular *smart distribution of computation* and related *advanced heterogeneous communication schemes*.

Both system's mission tasks and decision-making autonomy imply complexity for enabling *sensor fusion* from a multitude of integrated diverse sensors.

*Vehicle-to-vehicle communication* (V2V) has enabled efficient autonomy in complex dynamic networks of machines, e.g. heterogeneous swarms, at the price of very large overall systems' complexity.

Today, the industrial classification of autonomous systems' ICs in million gates (MG) assumes the following ranges: small-scale design <32MG; medium-scale <200MG; large-scale >200MG.

### C. Integrity

The main system integrity requirements brought to the front by autonomous systems use the *self-monitoring* and corresponding *self-awareness* that became a must and significantly support reliability enhancement. To cope with the diversity and complexity of environments the concepts such as *responsibility-sensitive safety* and *safety of the intended functionality* (SOTIF) [4] target at the autonomous system misuse cases and establish framework for responsibility sharing.

The challenge of high-severity failures in the system often dictates the requirement of the system to be *fail-safe* or even *fail-operational* thus also establishing diverse reliability requirements for modules to be available in the degraded mode. Here the complete system may benefit in some cases from the swarming of agents when the mission tasks can be re-allocated and accomplished even if some agent fails.

Finally, reliability validation is challenged by interference with requirements by other design aspects [20], [21] and correlation with security requirements in particular (e.g. [22] by ACEA association). For example, a new security standard under development (ISO 21434) is aiming at defining a Cybersecurity Assurance Level (CAL), similar to the ASIL concept [13].

## VI. CHALLENGES BY NOVEL HW ARCHITECTURES

In most cases, the requirements specified for autonomous systems cannot be fulfilled without resorting to advanced computing architectures and semiconductor technologies.

Concerning the architectures, this means that conventional CPU-based ones are substituted by alternative ones, including not only multicore devices, but also special modules, such as *General-Purpose Graphic Processing Units* (GPGPUs). Since Artificial Intelligence is widely used in autonomous systems, accelerators specifically targeting neural networks, such as the *Tensor Processing Units*, or TPUs, by Google, are intensively investigated and increasingly adopted. Complex innovative architectures such as the massively parallel, low-precision floating-point compute *Intelligence Processing Unit* (IPU) [25] require the smallest technology nodes to allow the necessary density of transistors bringing along the atomic scale reliability challenges. Finally, for some applications it may be convenient to include in the architecture some *FPGA modules* able to provide the flexibility to dynamically change the hardware supporting the implemented functions. All the mentioned components may be used as single devices or integrated into even more complex Systems on Chip (SoCs).

Moving from conventional architectures to the mentioned ones rises several important issues, not only in terms of hardware design complexity (e.g. in terms of validation) and software development and qualification, but also in terms of getting a sufficient understanding of their sensitivity to possible faults.

While several theoretical and experimental analysis provided information about the *Architecture Vulnerability Factor* [26] of traditional architectures [27], only preliminary results have been provided concerning the new ones and about the main modules they are composed of [28],[29],[30]. Moreover, these analyses are currently made more complex by the difficulty in getting representative open source models for the new architectures [31]. Similarly, the impact of possible faults affecting the new architectures when executing some common kinds of applications used in autonomous systems (e.g. those related to video processing and to neural networks implementation) is only partly understood [32].

On the other side, the requirements in terms of complexity, speed, power consumption and miniaturization force the adoption of advanced semiconductor technologies. Even if we stick to CMOS technologies, those that are going to be used for autonomous systems are widely unknown in terms of reliability. Hence, their adoption in safety-critical applications (as those of autonomous vehicles often are) asks first for effectively developing *new fault models*, given that the currently adopted ones are largely unsuitable to deal with the new defects characterizing these technologies. Secondly, *solutions* to detect and possibly tolerate these faults should be identified, taking into account that they should be able to trade-off their effectiveness with several other parameters, including cost and time-to-market. Finally, even when solutions will be available, their adoption will be possible only if *EDA tools* supporting them in a fully integrated manner with respect to the design and test flow will exist.

All the above issues are likely to become even more critical when non Von Neumann architectures and post-CMOS technologies will start to be adopted [33].

## VII. Reliability Enhancement

This Section aims at briefly summarizing some crucial points related to the current status of the art in the area of solutions able to achieve the required level of reliability when the electronic part of autonomous systems is considered.

As explained in the previous Sections, ASs are likely to expand significantly in the next years, provided that some of the technical and organization issues we are summarizing will be successfully overcome. Since now, we can imagine some trends that may be followed in the next future, based on what is happening in some representative and more advanced domains within the wide area of ASs, such as the automotive one.

Since the adopted technologies are intrinsically less reliable, a first trend goes into the direction of developing *solutions at the architecture level* that may guarantee by construction the target level of reliability. Solutions based on Duplication With Comparison (DWG), such as lockstep, are increasingly adopted when fault detection is the main target [34]. Unfortunately, their extension to the new architectures described in Section VI is not straightforward, although similar solutions implemented at the software level have been successfully explored already, especially with regular structures such as those of GPGPUs. Clearly, the adoption of hardware DWG architectures requires the development of specific products targeting safety-critical applications, only. Some recent products (e.g. Xavier by NVIDIA) go in this direction. Other products (e.g. the solution named Split-Lock introduced by ARM), although based on more conventional architectures, allow the user to dynamically decide whether to use the available redundancy to increase performance or reliability.

Given the high cost of solutions able to tolerate faults resorting to hardware redundancy (e.g., based on Triple Modular Redundancy, or TMR), alternative solutions exploiting *reconfiguration* seem particularly attractive [35],[36]. In particular, they may provide a mechanism to extend the lifetime of adopted circuits, whose span is quickly shrinking in advanced technologies, and tends to be increasingly dependent on the operating environment and workload [37]. Additionally, the new technologies may increase the chance that multiple faults occur in a logic block, as it already happens for memory ones, and this may increase the negative impact of fault accumulation, which can hardly be managed via TMR. Reconfiguration can be applied at different levels and managed either directly in hardware, or resorting to Operating System features. In all cases, suitable techniques able to quickly detect (and possibly locate) faults are crucial. Existing *Design for Testability structures* (e.g. supporting Logic BIST for chunks of logic, or Memory BIST for memory modules) already introduced to support end-of-manufacturing test may be re-used for this purpose. In other cases users resort to functional solutions, e.g. based on the so-called Self-Test Libraries [38], which rely on the Software-based Self-test approach [39],[40]. This solution seems particularly effective for in-field test of complex systems, since it allows to exploit self-test code provided by the developer of the modules composing the system and able to achieve a given fault coverage, which is integrated by the user into the application code and run when required to achieve the target reliability figures (e.g., at the Power-On, or periodically, or when specific error conditions happen). The results produced by the system when executing such pieces of code allow the detection of possible faults [41]. Preliminary results show that the same approach can be extended also to new architectures, such as GPGPUs [42].

Another interesting research direction which is proving to be promising for autonomous systems lies in *trading-off precision with reliability*. Preliminary results about techniques where hardware resources saved by moving to a lower precision are used to increase reliability are shown in [43].

Given the complexity of the systems, we are targeting, in most cases they will integrate components designed and produced by different companies. Hence, cross-layer approaches to reliability are highly promising [44], but require a clear definition of what each level should guarantee in terms of reliability, and how this can be validated.

It is worth mentioning that ASs are characterized by a wide variety of scenarios and constraints. In some of them, different and highly independent systems will *cooperate* to achieve a given target (Systems of Systems, or SoSs). In such cases, new paradigms may emerge even from the point of view of reliability. The complexity and heterogeneity of the resulting SoS may favor the adoption of holistic solutions which will enable it to implement highly innovative features, such as self-reconfiguration, self-test, and implicit robustness.

As a final comment, we would like to emphasize the already mentioned *increasing importance of security* for autonomous systems [5]. Several works highlighted that facing security often requires adopting solutions based on opposite strategies with respect to those required by reliability and test, e.g. in terms of system status observability. For this reason, integrated solutions identifying suitable trade-off between opposite constraints are required [45].

VIII. CONCLUSIONS

The paper has presented an overview of reliability challenges related to the novel and very rapidly developing domain of autonomous systems. The challenges of reliability assessment and enhancement stem from a set of general attributes, novel applications' specific requirements and new hardware architectures of autonomous systems. The way forward for the research community and industry lays in understanding the new needs, collaboration towards comprehensive solutions and adoption of new appropriate standards.

ACKNOWLEDGMENTS

This research was supported in part by projects H2020 MSCA ITN RESCUE funded from the EU H2020 programme under the MSC grant agreement No.722325, by the Estonian Ministry of Education and Research institutional research grant IUT19-1 and by European Union through the European Structural and Regional Development Funds.